\documentclass[preprint,showpacs,preprintnumbers,amsmath,amssymb,aps]{revtex4-1}

\usepackage{graphics}
\usepackage{longtable} 
\usepackage{subfigure}
\usepackage[normalem]{ulem}
\usepackage{wrapfig}
\usepackage{epsfig}
\usepackage{float}
\usepackage{verbatim}
\usepackage{hyperref}
\usepackage{amsmath,amssymb,bm}	
\usepackage{fancyheadings}
\usepackage{graphicx}
\usepackage{array}
\usepackage{psfrag}
\usepackage{color}
\usepackage{bbold}
\usepackage{ulem}
\begin{document}

\title{A Critical Assessment of the Boltzmann Approach for Active Systems}

\preprint{LMU-ASC 74/13}

\author{Florian Th\"uroff,* Christoph A. Weber,* and Erwin Frey}

\affiliation{Arnold Sommerfeld Center for Theoretical Physics and Center for NanoScience, Department of Physics, Ludwig-Maximilians-Universit\"at M\"unchen, Theresienstrasse 37, D-80333 Munich, Germany}
\altaffiliation{F. Th\"uroff and C.A. Weber contributed equally to this work.}

\begin{abstract}
Generic models of propelled particle systems posit that the emergence of polar order is driven by the competition between local alignment and noise. Although this notion has been confirmed employing the Boltzmann equation, the range of applicability of this equation remains elusive. We introduce a broad class of mesoscopic collision rules and analyze the prerequisites for the emergence of polar order in the framework of kinetic theory. Our findings suggest that a Boltzmann approach is appropriate for weakly aligning systems but is incompatible with experiments on cluster forming systems.
\end{abstract}

\maketitle

The emergence of large-scale collective motion is one of the most intriguing features, which is shared among a large variety of active systems~\cite{Butt, Schaller, schaller2, Yutaka,Dombrowski_2004, pnas_bacteria,Dauchot_Chate_2010, Dauchot_long, Kudrolli_2008,ballerini_starlings,Ringe_PNAS}. This  apparent universality
has led to considerable theoretical efforts aiming at the identification of general physical principles underlying collective motion in active systems~\cite{Aranson_Tsimring_book, Julicher_gel_review, Vicsek_Review,  Ramaswamy_Review, Marchetti:2012ws}. There is a general consensus that an antagonism between dynamic processes favoring alignment between the particles' velocities and noise is the basic mechanism which triggers a phase transition from an isotropic to a polar ordered state. The computational model by Vicsek et al.~\cite{Vicsek} adapts this idea and implements it as an update rule in the spirit of a cellular automaton: Each particle aligns parallel to the average of all particles' orientations within some finite neighborhood. 

\begin{figure}[tb]
	\centering
		\includegraphics[width=.4\textwidth]{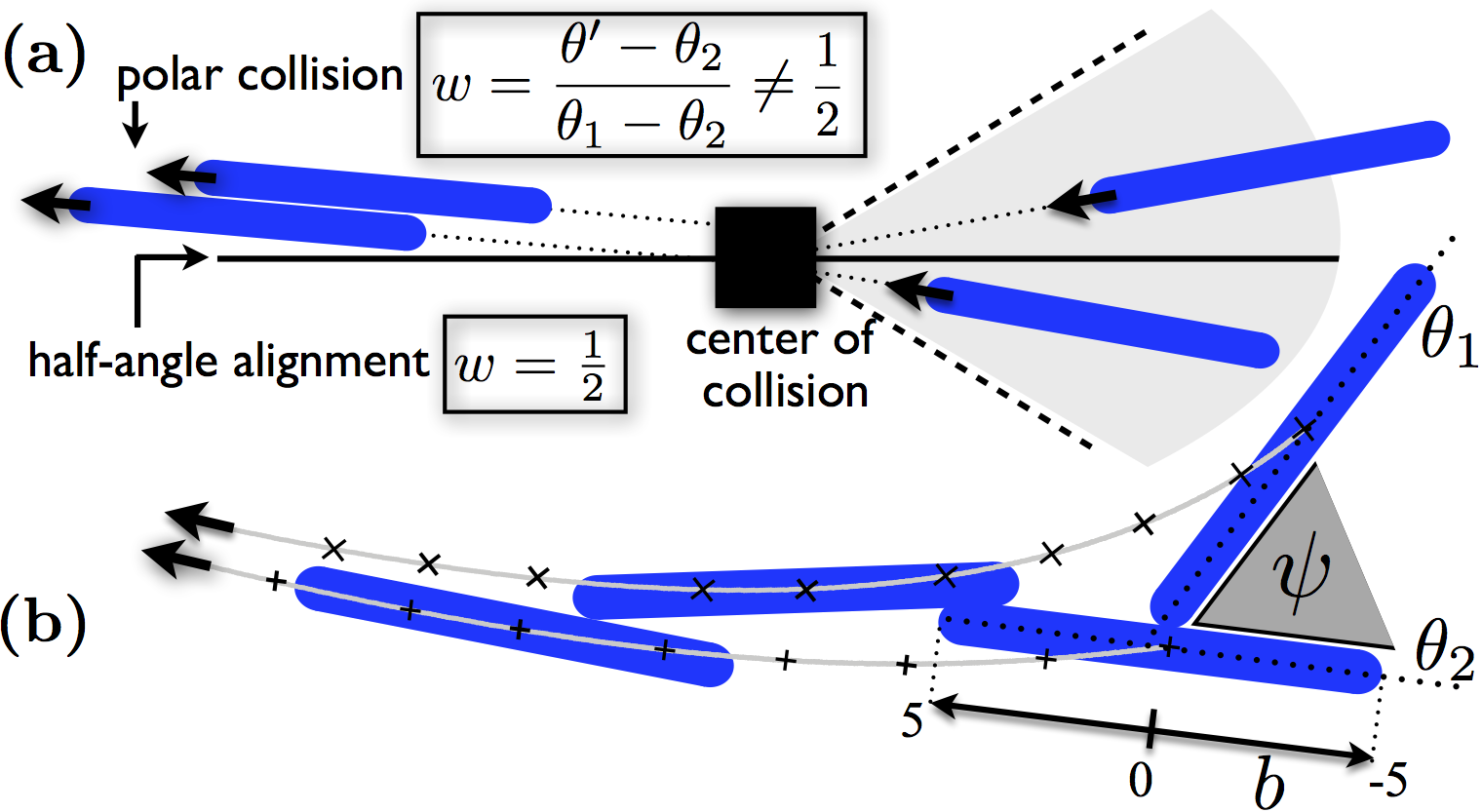}
	\caption{\textbf{(a)} Rod interactions within an angular alignment range ${\bar \psi}$ (grey shaded) lead  to polar alignment, whose direction typically deviates from the half-angle alignment $w=\frac{1}{2}$ [$w$ is directly given by Eq.~\eqref{eq:MesoCollision}]. \textbf{(b)} Simulated example trajectory for a model of propelled rigid rods  (relative angle $\psi\approx\pi/3$, impact parameter $b\approx0$ in units of rod diameter, $w(\pi/3,0)\approx1.3$).
	}
	\label{fig:amean}
\end{figure}

On a more microscopic scale, Bertin et al.~\cite{Bertin_short,*Bertin_long} have formulated the dynamics of propelled particle systems in the framework of kinetic theory. Collisions between particles are described by a half-angle alignment rule, \emph{i.e.}\ one assumes that particles line up parallel to their average orientation upon binary collisions. While, similar to the Vicsek model, this rule phenomenologically accounts for the dissipative character of collisions, it is not rooted in a microscopic analysis of actual collision processes between active particles. In the meantime there are, however, a range of well-characterized experimental model systems including actin and microtubule gliding assays~\cite{Schaller, Yutaka}, and shaken granular particles~\cite{Dauchot_long, VibratedDisks2013}, which are amenable to a quantitative analysis at the scale of collisions between individual particles. These studies highlight that actual collisions differ from generic interaction rules in two important respects: (i) The post-collision particle orientations are not symmetric with respect to the average of the pre-collision directions, but depend on both the relative orientation and relative position of the colliding particles before the collision [Fig.~\ref{fig:amean}]. (ii) Frequently, one finds `indifferent' collision events where collisions do not change the relative orientations of the collision partners~\cite{Dauchot_long, VibratedDisks2013}.  Therefore, the region of configuration space supporting aligning collisions is restricted, and it is thus far from obvious whether binary, dissipative interactions actually contribute to the formation of order in active systems. This raises an important question: With no additional assumptions to be made, does the integrated effect of binary particle interactions in active systems suffice to establish a state of collective motion on macroscopic scales? 

To address this question, we investigate the physics of binary collisions between propelled rod-like particles and its impact on the emergence of polar order within the framework of kinetic theory. Our starting point is a numerical analysis of the collision process between propelled rod-like particles. The goal is to motivate a broad class of mesoscopic collision rules comprising the most pertinent features of the dissipative collisions. 
Integrating this collision rule into a Boltzmann equation, we give a systematic study of the ordering capabilities of active systems in terms of 
the underlying collision dynamics.
Our findings suggest that a kinetic description based on binary particle interactions is suitable to capture 
ordering processes which proceed via a gradual reduction in the spread of particle orientations, but fails if collective motion patterns emerge from clustering processes. 
Within our kinetic framework, we demonstrate that the presence of noise in the collision process (collision noise) imposes
a `minimum efficiency requirement' on the underlying microscopic collision dynamics if collective motion is to be observed. Finally, we  address the system's spatio-temporal behavior using a numerical solution of the underlying kinetic equations~\cite{Thuroff_2013}. We demonstrate that, for a large class of generic collision rules, the formation of wave-like patterns, as previously observed in agent-based simulations (see, e.g., Ref.~\cite{Chate_long}), is a robust feature accompanying the transition to collective motion, and rendering this phase transition discontinuous.

\begin{figure}[tb]
	\centering
		\includegraphics[width=.5\textwidth]{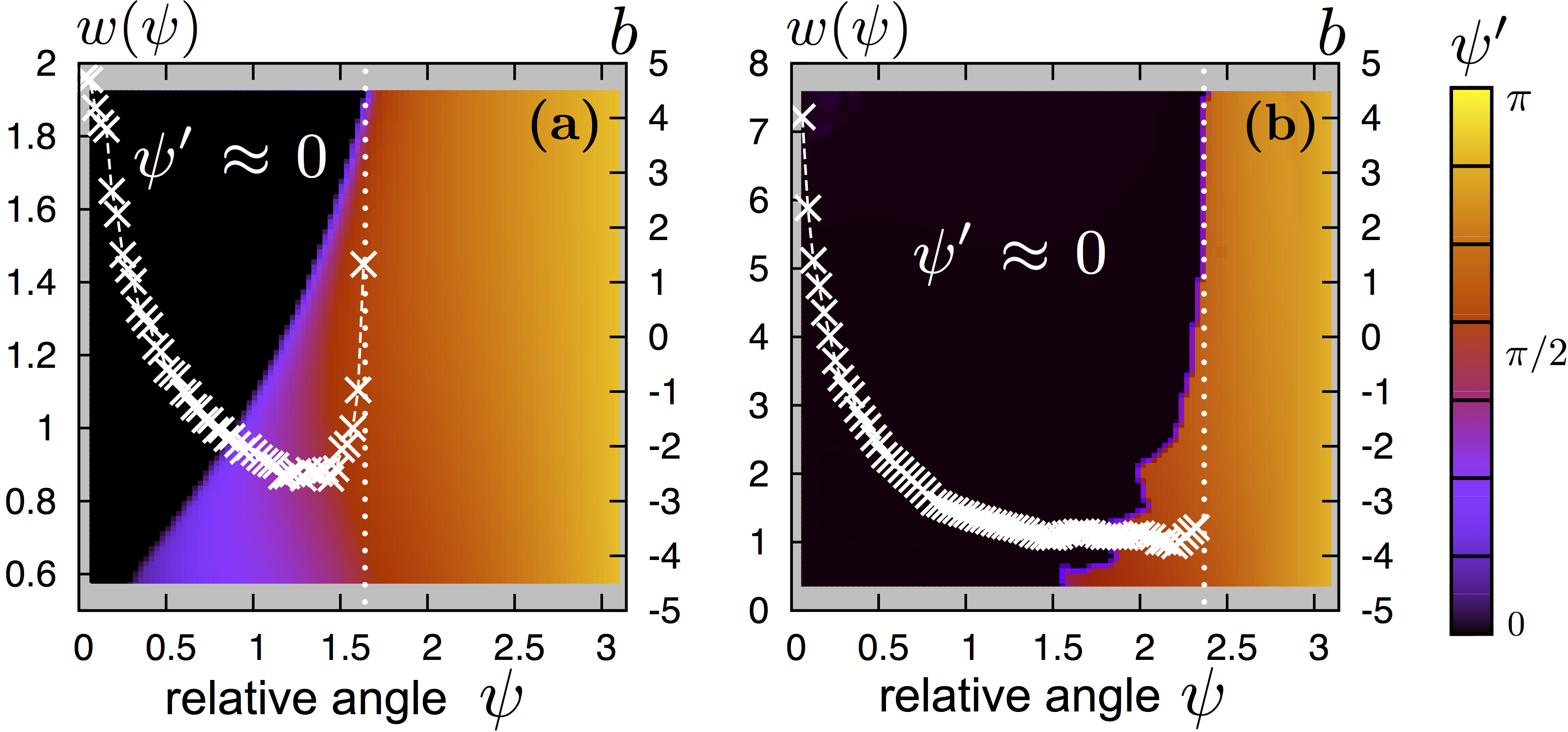}
	\caption{\emph{Binary scattering plot:} relative post-collision angle $\psi'=\theta_1'-\theta_2'$ (scale bar), versus impact parameter $b$ and relative pre-collision angle $\psi$, for propelled hard rods \textbf{(a)}, and stiff polymers \textbf{(b)}. The mesoscopic alignment weight $w(\psi)$, and the alignment range ${\bar \psi}$ are indicated as crosses and vertical dotted lines, respectively. $L= 10$; for simulation details and videos refer to Supplementary Material~\cite{Supplement}.}
	\label{fig:Scatter_plot}
\end{figure}

\emph{Two microscopic scenarios.} To quantify binary particle collisions we performed numerical  simulations for two exemplary models of propelled rod-like particles (length $L=10$, diameter $d=1$) moving in a two-dimensional over-damped environment (for details see Supplementary Material~\cite{Supplement}). Specifically, we considered hard-core rigid, propelled rods  as in~\cite{Peruani_rods}, and a \emph{bead-spring} model for propelled stiff polymers with a finite bending modulus, interacting by a short-ranged polar alignment interaction. Each particle is driven by a constant propelling force, pointing along the long axis of the rod or parallel to the polymer's contour. We performed scattering studies by preparing the particles with different relative pre-collision orientations $\psi=\theta_1-\theta_2 \in[0, \pi]$ ($\theta_{1/2}$: pre-collision rod orientations) and different impact parameters $b$ (measuring the location of impact along the ``target particle's'' contour; cf. Fig.~\ref{fig:amean}), 
and observe the resulting relative post-collision orientations $\psi'=\theta_1'-\theta_2'$ ($\theta_{1/2}'$: post-collision rod orientations). The corresponding results are summarized in Figs.~\ref{fig:Scatter_plot}(a,b).

Although quantitative differences are manifest, there are distinct qualitative features which equally apply to both scattering studies: Collisions may be classified either as \emph{`indifferent'}, where the relative scattering angle remains virtually unchanged ($\psi' \approx \psi$)~\footnote{Since indifferent events do, in general, change the mean orientation of the particles, they could be described by an effective noise term, which will be neglected here.}, or  \emph{`polar alignment'} events with $\theta_1'=\theta_2'=:\theta'$, \emph{i.e.}\ $\psi'$=0. These two regimes are delineated by a rather sharp boundary denoted as $\psi_{\text{max}}(b)$, which is the maximum value of $\psi$ in the regime where $\psi'\approx0$. For simplicity, we will approximate this boundary by a single angle ${\bar \psi}=\max_{b}\psi_\text{max}(b)$ which we term the \emph{effective alignment range} [indicated by vertical dotted lines in Fig.~\ref{fig:Scatter_plot}(a,b)]. Within the alignment range ($\psi \leq {\bar \psi}$), the post-collision angle $\theta'$ is an intricate function of the pre-collision angles and the impact parameter. Due to rotational invariance it is of the form $\theta'= w(\psi,b)\,\theta_1 + [1-w(\psi,b)]\,\theta_2$, where $w(\psi,b)$ can be interpreted as \emph{microscopic alignment weight} characteristic for the respective model; cf. Fig. \ref{fig:amean}(a). Since we are aiming at a Boltzmann approach which does not resolve length scales comparable to the size of single particles, we adopt a mean-field approach and average over all impact parameters to introduce a \emph{mesoscopic alignment weight} $w(\psi) := \langle w(\psi,b)\rangle_b$. It defines the relative magnitude of the pre-collision angles in the post-collision angle:
\begin{equation}
\theta' = w(\psi)\,\theta_1 + [1-w(\psi)]\,\theta_2.
\label{eq:MesoCollision}
\end{equation}
Taken together, the alignment range ${\bar \psi}$ and the alignment weight $w(\psi)$ constitute a \emph{generalized mesoscopic collision rule}. It accounts for indifferent scattering events as well as deviations from half-angle alignment ($w=\frac12$). Fig.~\ref{fig:Scatter_plot} shows the actual form of both quantities for the particular examples discussed above. Next, abstracting from the two particular examples given, we explore the consequences of such generalized collision rules for the ordering propensity of active matter.

\emph{Analytical formulation.} The Boltzmann equation for driven particles in two spatial dimensions takes the following general form~\cite{Bertin_short,*Bertin_long}:
\begin{subequations}
\label{eq:BoltzmannEquation}
\begin{equation}
\partial_t f(\vec{r},\theta,t)+v_0{\vec{e}}(\theta)\cdot\nabla f(\vec{r},\theta,t)=I_{\text{d}}+I_{\text{c}}.
\end{equation}
Here $f(\vec{r},\theta,t)$ is the one-particle distribution function, and $v_0$ and ${\vec{e}}(\theta)$ denote the magnitude and direction of the velocity vector of freely moving particles, respectively. $I_d$ accounts for rotational particle diffusion with Gaussian-distributed angular increments $\varphi$ [with standard deviation $\sigma_0$] occurring with a rate $\lambda$~\footnote{The argument of $f(\phi)$ is understood modulo $2\pi$.}  
\begin{equation}
I_{\text{d}} = \lambda\,  \Bigr\langle \int_{-\pi}^{\pi}d\phi f(\phi) [\delta(\theta-\phi-\varphi)-\delta(\theta-\phi)]  \Bigl\rangle_{\varphi}.
\label{eq:DifusionIntegral}
\end{equation}
The collision integral 
\begin{eqnarray}
\nonumber
I_{\text{c}} &=& \Bigr\langle \int_{-\pi}^{\pi}d\phi\int_{-{\bar \psi}}^{{\bar \psi}}d\psi\,\mathcal{S}(L,d,\psi)f(\phi)f(\phi+\psi)\\
\label{eq:CollisionIntegral}
&&\times\left[\delta(\theta-\phi-w(\psi)\psi-\eta)-\delta(\theta-\phi)\right] \Bigl\rangle_{\eta},
\end{eqnarray}
\end{subequations}
quantifies binary scattering events and depends on the mesoscopic collision rule. In Eq.~\eqref{eq:CollisionIntegral},  $\mathcal{S}(L,d,\psi)= 4v_0d\, \left|   \sin\left(\frac{\psi}{2}\right)\right|\left(1+\frac{L/d-1}{2}\left|\sin\psi\right|\right)$
 denotes the ``differential scattering cross section''~\cite{Weber_NJP_2013}. We have also included stochastic effects by adding a Gaussian-distributed angle $\eta$ [with standard deviation $\sigma$] to the post-collision angle.

To calculate the onset of polar order we analyze Eq.~\eqref{eq:BoltzmannEquation} in terms of Fourier modes $\hat{f}_k = \int_{-\pi}^{\pi}d\theta\,e^{i k \theta}f(\theta)$:
\begin{equation}
\label{eq:BoltzmannFourierSpace}
\begin{split}
&\partial_t\hat{f}_k+\frac{v_0}{2}\left[\partial_x(\hat{f}_{k+1}+\hat{f}_{k-1})-i\partial_y(\hat{f}_{k+1}-{\hat{f}}_{k-1})\right]=\\
&-\lambda\left(1-e^{-(k\sigma_0)^2/2}\right){\hat{f}}_k+\sum_{n=-\infty}^{\infty}\mathcal{I}_{n,k}{\hat{f}}_n{\hat{f}}_{k-n},
\end{split}
\end{equation}
where the Fourier coefficients of the collision kernel $\mathcal{I}_{n,k}$ are given in~\cite{Supplement}. The first two Fourier components, are related to the (hydrodynamic) particle density $\rho={\hat{f}}_0$ and the momentum density $\vec \tau=\rho\vec u=v_0{\hat{f}}_1$, where $\vec u$ denotes the hydrodynamic velocity. We are interested in the regime close to the transition from an isotropic state, $f(\vec r,\theta,t)=\frac{1}{2\pi}$, toward a polar state.
For our base states we then assume $|\vec u|  \ll v_0$ and, consequently, $|{\hat{f}}_k|=\mathcal{O}\left[(|\vec u|/v_0)^k\right]\ll1$.
We are thus able to truncate~Eqs.~\eqref{eq:BoltzmannFourierSpace} by setting ${\hat{f}}_k\approx0$ for all  $k>2$~\cite{Aronson_MT,Bertin_short,*Bertin_long}. Since the instability of the isotropic state toward formation of collective order occurs at zero wave number~\cite{Supplement}, we may restrict ourselves to spatially homogeneous systems and drop all spatial derivatives: 
\begin{subequations}
\label{eq:HydrodynamicEquations}
\begin{eqnarray}
\label{eq:HydrodynamicEquationsRho}
\partial_t\rho&=&0,\\
\label{eq:HydrodynamicEquationsTau}
\partial_t\bm{\tau}&=&-\nu_1\,\bm{\tau}+\nu_2^{-1}\left[\mathcal{I}_{1,2}(\mathcal{I}_{-1,1}+\mathcal{I}_{2,1})\right]\bm{\tau}^2\,\bm{\tau},\quad
\end{eqnarray}
\end{subequations}
where $ \nu_k = 1-e^{-(k\sigma_0)^2/2}-\rho\left(\mathcal{I}_{0,k}+\mathcal{I}_{k,k}\right)$.
In Eqs.~\eqref{eq:HydrodynamicEquations} variables have been rescaled: $t\rightarrow t/\lambda$, $\rho\rightarrow\rho \lambda/\bar{\mathcal{S}}$, $\bm{\tau}\rightarrow\bm{\tau}v_0\lambda/\bar{\mathcal{S}}$, $\mathcal{I}_{n,k}\rightarrow\mathcal{I}_{n,k}\bar{\mathcal{S}}$ with $\bar{\mathcal{S}}=(2\pi)^{-1}\int_{-{\bar \psi}}^{{\bar \psi}}\mathcal{S}(L,d,\psi)d\psi$ the ``total scattering cross section'' for polar collisions. In these units, $\rho$ measures the frequency of collisions relative to the frequency of self-diffusion, $\lambda$.
Depending on the alignment range ${\bar \psi}$, $\nu_2$ might turn negative, in which case Eqs.~\eqref{eq:HydrodynamicEquations} would have to be complemented by equations for higher order broken-symmetry variables, like the nematic tensor~\cite{Peshkov:2012tu}. Here we restrict ourselves to the discussion of the polar case, where $\nu_2>0$. Then, Eq.~\eqref{eq:HydrodynamicEquationsRho} simply expresses conservation of particle number, and Eq.~\eqref{eq:HydrodynamicEquationsTau} captures the formation of collective motion via spontaneous breaking of rotational symmetry. 

The isotropic state $\bm\tau=0$ becomes unstable at a threshold density $\rho_t$, determined by $\nu_1 (\rho_t) = 0$:
\begin{equation}
\label{eq:rhoc}
\rho_t[w(\psi);{\bar \psi}]=\frac{1-e^{-\sigma_0^2/2}}{\mathcal{I}_{0,1}[w(\psi);{\bar \psi}]+\mathcal{I}_{1,1}[w(\psi);{\bar \psi}]} \, .
\end{equation}
Importantly, the analytical form of $\rho_t$, Eq. \eqref{eq:rhoc}, is exact, \emph{i.e.}\ independent of the particular scheme used to truncate the Fourier space Boltzmann equation~\eqref{eq:BoltzmannFourierSpace}, and establishes a direct connection between the microscopic details of particle collisions and the location of the phase boundary toward collective motion. Subsequent quantitative investigations of this connection are greatly facilitated by the fact that the functional space of all alignment weight functions $w(\psi)$ can be organized in terms of equivalence classes, using the equivalence relation $w_1(\psi)\sim w_2(\psi)$ if and only if $\rho_t[w_1(\psi)]=\rho_t[w_2(\psi)]$. It can be shown \cite{Supplement} that each such equivalence class contains one and only one constant alignment weight function $\bar w$, which can be determined from any (e.g. experimentally measured) $w(\psi)$ by inverting the relation $\rho_t[w(\psi)]=\rho_t[\bar w]$. In what follows, we will refer to $\bar w$ as \emph{``effective alignment weight''} and use it as a convenient ``parameterization'' of the functional space of all alignment weight functions. The symmetry of the Boltzmann equation~\eqref{eq:BoltzmannEquation} with respect to ${\bar w}\rightarrow1-{\bar w}$, allows to consider ${\bar w} \geq 0.5$ without loss of generality.

\begin{figure}[tr]
	\centering
		\includegraphics[width=.5\textwidth]{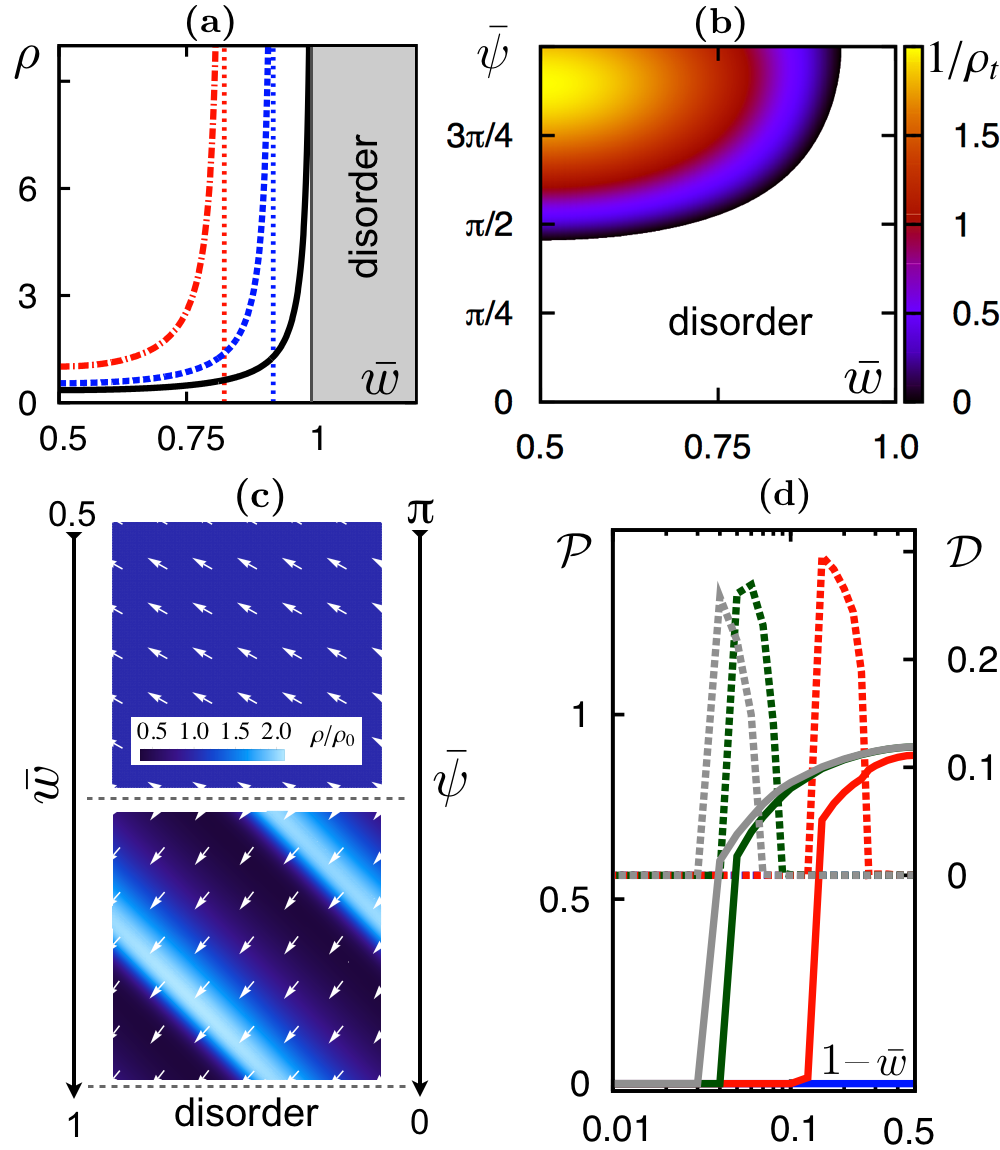}
	\caption{(color online) 
	 \textbf{(a)} \textit{Transition density $\rho_t(\bar w)$}: 
	 Curves from left to right (dash-dotted, dashed, solid) correspond to $\sigma=\{0.7, 0.5, 0 \}$ with ${\bar \psi}=\pi$.
 \textbf{(b)}  $\rho_t^{-1}({\bar w}, {\bar \psi})$ for $\sigma=0.5$.
Parameters for \textbf{(a)} and \textbf{(b)}: $L/d=1$, $\sigma_0=0.5$.
	  \textit{Stationary numerical solution of Eq.~\eqref{eq:BoltzmannEquation}}: 
	    \textbf{(c)} Snapshots of observed polar patterns (periodic boundary conditions; arrows: normalized, local polarization) for ${\bar \psi}=3\pi/4$: homogeneously distributed density field (top, $\bar w=0.8$) and polar-wave pattern (bottom, $\bar w=0.92$, relative density differences are indicated, see color bar). 
	    \textbf{(d)} Total polarity $\mathcal{P}$ (solid lines) and spatial density variance $\mathcal{D}$ (dashed lines) versus ${\bar w}$ (from left to right: grey-green-red-blue correspond to ${\bar \psi}=\{\pi, 3\pi/4, \pi/2, \pi/4\}$). For $\psi=\pi/4$ (blue), $\mathcal{D}=0$ and $\mathcal{P}=0$  for all $\bar w$. 
	    Parameters for \textbf{(c)} and \textbf{(d)}: $\rho_0=0.15$, $\sigma=\sigma_0=0.15$, $L/d=1$.
	  }
	\label{fig:Homogeneous_analytics}
\end{figure}

\emph{Analytical results.} Fig.~\ref{fig:Homogeneous_analytics}(a) shows the threshold density $\rho_t$ as a function of the \emph{effective alignment weight} ${\bar w}$ and different noise strengths $\sigma$.
 
For deterministic collisions, $\sigma = 0$, the transition density $\rho_t$ is finite over the range ${\bar w} < 1$, and diverges
as ${\bar w} \to 1$ [solid curve in Fig.~\ref{fig:Homogeneous_analytics}(a)]. In the latter limit, the post-collision angle coincides with one of the pre-collision angles, cf.~Eq.~\eqref{eq:MesoCollision}. This type of binary collision is on average equivalent to a specular reflection: Due to particle exchange symmetry, two particles with pre-collision angles $\theta_1$ and $\theta_2$ are both  either scattered into $\theta_1$ ($\nearrow \nwarrow \, \, \rightarrow \, \, \nwarrow\nwarrow$) or $\theta_2$ ($\nearrow \nwarrow \, \, \rightarrow \, \, \nearrow\nearrow$), 
with equal probability. 
Since 
two-particle correlations are not taken into account in the Boltzmann equation, 
we have $\frac{1}{2}\left(\nwarrow\nwarrow+\nearrow\nearrow \right)\equiv\, \, \nwarrow \nearrow$, 
which amounts to a specular reflection.
This \emph{``alignment--reflection duality''} carries over to arbitrary collision processes ($\bar w \neq 1$), and is inextricably linked to the Boltzmann equation due to the molecular chaos assumption.
As a consequence, polar order cannot build up from a disordered state 
for effective alignment weights $\bar w \geq 1$.
In the light of this discussion, the parameter ${\bar w} - 1\equiv\delta$ can be reinterpreted as an \emph{angular dispersion factor}: While for $\delta =0$ the angular distribution remains invariant,  any deviation from an isotropic distribution is amplified by collisions for $\delta <0$
and polar order develops.
\emph{Angular dispersion factors} $\delta > 0$ (\emph{i.e.} $\bar w>1$) have the opposite effect. Obviously, the threshold density $\rho_t$ also diverges for ${\bar \psi}\to0$ (data not shown), which corresponds to the limiting case where all collisions are indifferent, \emph{i.e.}\ to non-interacting particles.

The above picture is to be modified upon adding stochasticity to the collisions, $\sigma>0$. Then, the poles of $\rho_t$ are increasingly shifted toward the limiting case of optimal alignment conditions, \emph{i.e.}\ $({\bar w},{\bar \psi}) = (\frac12,\pi)$ [Figs.~\ref{fig:Homogeneous_analytics}(a,b)]. Surprisingly, to compensate for the dis-aligning effect of collision noise one needs both, a smaller angular dispersion $\delta$, and  a larger alignment range ${\bar \psi}$. Even for half-angle alignment, ${\bar w}=\frac{1}{2}$, any stochasticity during the collision process immediately sets a lower bound for the alignment range ${\bar \psi}$, which cannot be abrogated by increasing the density; cf. Fig.~\ref{fig:Homogeneous_analytics}(b). In other words, the presence of collision noise imposes a ``minimum efficiency requirement'' on microscopic particle interactions for the emergence of polar order.

The popular half-angle alignment rule thus overestimates the effect of binary collisions on the build-up of orientational order. This is indeed the case for the two paradigmatic scenarios of propelled rods and stiff polymers discussed above: Using the microscopic scattering data $\psi_{\text{max}}(b)$ and $w(\psi,b)$ from our simulations, we computed the transition density $\rho_t$ by means of  Eq.~\eqref{eq:rhoc}. To get a precise estimate we explicitly accounted for the functional dependence of the collision integrals on the \emph{microscopic} alignment weight $w(\psi, b)$, only assuming uniformly distributed collision parameters $b$. Interestingly, the ensuing threshold densities for both models turn out to be negative. The failure of these models to establish a state of polar order suggests to reconsider the range of applicability of the Boltzmann approach to active systems.
In fact, recent numerical and experimental work on similar systems \cite{Peruani_rods,Weber_nucleation,Peruani:2012wc} highlights the importance of nucleation and growth of clusters as a driving force behind the formation of large scale non-isotropic structures. Within these clusters, however, correlation effects become important which are incompatible with the \emph{alignment--reflection duality} alluded to above. Thus, while the Boltzmann approach is well suited to capture the emergence of order via a gradual reduction in the spread of particle orientations, it breaks down if ordering proceeds via the formation of clusters.

\emph{Numerical solution of the Boltzmann equation.} To address the impact of ${\bar w}$ and ${\bar \psi}$ on the system's capability to form patterns, we numerically solved the Boltzmann equation~\eqref{eq:BoltzmannEquation}  using a constant alignment weight, $w(\psi)={\bar w}$~\cite{Supplement,Thuroff_2013}.
We tested that the numerical solution reproduces the phase boundary obtained from our analytical calculations, Eq.~\eqref{eq:rhoc} (data not shown).
In accordance with previous agent-based simulations~\cite{Chate_long} and with expectations from analytical considerations \cite{Bertin_long}, we observe stable traveling wave patterns
emerging from random initial conditions
for $\rho \gtrsim \rho_t({\bar w},{\bar \psi})$, and with typical asymmetrical wave front profiles; cf.~snapshot Fig.~\ref{fig:Homogeneous_analytics}(c) and videos in the Supplemental Material~\cite{Supplement}.
Fig.~\ref{fig:Homogeneous_analytics}(d) illustrates the onset of ordering upon varying the \emph{effective alignment weight} $\bar w$ at a fixed overall density $\rho_0>\rho_t[\bar w=0.5,\bar\psi=\pi]$, fixed noise parameters, and a set of \emph{alignment ranges $\bar\psi$}:
Crossing the transition point (at $\rho_t[\bar w,\bar\psi]=\rho_0$), we observe a growth of the system's total polarity $\mathcal{P}=\overline{\bm{\tau}(\vec r)}$ [overbar: spatial average], which is accompanied by the formation of traveling wave patterns in a parameter window adjacent to the ordering transition. As a simple indicator for wave-like patterns we use the variance of the spatial density $\mathcal{D}=\overline{(\rho(\vec r)-\overline{\rho})^2/\overline{\rho}^2}$ measured at times where the traveling wave patterns become stationary. The emergence of wave-like patterns in the transition region toward polar order seems to be a generic consequence of linear collision rules, \emph{i.e.}\ constant \emph{alignment weight functions} $w(\psi) = \bar w = \text{const}$.

Contrary to what can be expected from the spatially homogeneous dynamics, Eq. \eqref{eq:HydrodynamicEquationsTau}, our numerical solutions [Fig. \ref{fig:Homogeneous_analytics}(d)] indicate that the development of spatial inhomogeneities in the form of traveling wave fronts causes the polar order parameter $\mathcal{P}$ to undergo a discontinuous jump. This indicates a first-order phase transition toward polar order, rather than a second-order transition. An analogous transition behavior is observed upon varying the density $\rho$ at constant values for the collision parameters (not shown).
Whether or not a more complicated, non-linear collision dynamics alters this behavior poses an interesting question for future work.

\begin{acknowledgments}
We acknowledge support by the Deutsche Forschungsgemeinschaft in the framework of the SFB 863 ``Forces in Biomolecular Systems'', and the German Excellence Initiatives via the program ``NanoSystems Initiative Munich (NIM)''. This research was also supported in part by the National Science Foundation under Grant No. NSF PHY11-25915.
\end{acknowledgments}

\end{document}